%% file: main.tex
\documentclass[submission,copyright,creativecommons]{eptcs}
\providecommand{\event}{ALP4IoT 2016} % Name of the event you are submitting to
\usepackage{underscore}           % Only needed if you use pdflatex.

\usepackage{amsmath,amsthm}
\usepackage{graphicx}
\usepackage[usenames,dvipsnames,svgnames,table]{xcolor}
\usepackage{stmaryrd}
\usepackage{fancyvrb}
\usepackage{lipsum}

\input{macros}

\usepackage{setspace}
\usepackage{titlesec}
\titlespacing*{\section}{0pt}{3.5ex plus 1ex minus 2ex}{2.3ex plus .2ex minus 1ex}
\title{Resilient Blocks for Summarising Distributed Data}
\author{Giorgio Audrito
\institute{University of Torino, Italy}
\email{giorgio.audrito@unito.it}
\and
Sergio Bergamini
\institute{University of Torino, Italy}
\email{sergio.bergamini@edu.unito.it}
}
\def\titlerunning{Resilient Blocks for Summarising Distributed Data}
\def\authorrunning{G. Audrito \& S. Bergamini}

\begin{document}
\maketitle

\begin{abstract}
	Summarising distributed data is a central routine for parallel programming, lying at the core of widely used frameworks such as the \emph{map/reduce} paradigm. In the IoT context it is even more crucial, being a privileged mean to allow long-range interactions: in fact, summarising is needed to avoid data explosion in each computational unit.

	We introduce a new algorithm for dynamic summarising of distributed data, \emph{weighted multi-path}, improving over the state-of-the-art \emph{multi-path} algorithm. We validate the new algorithm in an archetypal scenario, taking into account sources of volatility of many sorts and comparing it to other existing implementations. We thus show that \emph{weighted multi-path} retains adequate accuracy even in high-variability scenarios where the other algorithms are diverging significantly from the correct values.
\end{abstract}

\section{Introduction}

% % why care about spatial computing? / spatial computing through composition of building blocks

The modern world is increasingly permeated by heterogeneous connected devices, establishing a fully-integrated digital and physical ecosystem. Such a pervasive landscape calls for the adoption of self-organising mechanisms, for which it is challenging and critical to translate the system global requirements into single-agent behaviors. In this context, each agent has a limited knowledge of its environment, which has to be shared with neighbors in order to allow for coordination mechanisms and create a global spatio-temporal representation of the distributed state of computation.
Several tools address this problem \cite{SpatialIGI2013}, showing how self-organization can be achieved using a small set of ``basic'' components, upon which increasingly complex algorithms and behaviors are built \cite{BPV-COMPUTER2015}. Due to the dynamic scenarios typical of pervasive computing, these components need to carefully trade-off efficiency with resilience to network changes.

% % building block C

The \emph{collection} building block (C in short) is one of the most basic and widely used components, which aggregates values held by devices in a spatial region into a single resulting value in a selected device. This procedure closely resembles the \emph{reduce} phase of the MapReduce paradigm \cite{google:mapreduce}, ported into a spatial computing context. This block can be applied to a variety of different contexts, as it can be instantiated for values of any data type with an associative and commutative aggregation operator. However, existing implementations of C (single-path and multi-path \cite{VBDP-SASO2015}) exhibit poor performance whenever node mobility is significant with respect to the device update rate.

% % paper contribution and structure

In this paper we introduce a new implementation for the C building block, called \emph{weighted multi-path}, which is able to achieve adequate accuracy in highly volatile scenarios. Its performance is then validated in a archetypal situation, taking into account node mobility, update rate variability and discontinuities in network configuration. Section \ref{sec:collection} presents the state-of-the-art implementations of C; Section \ref{sec:weighted} describes the new weighted multi-path algorithm; Section \ref{sec:evaluation} compares the new algorithm with other existing implementations in a archetypal scenario; Section \ref{sec:conclusion} summarises the contributions of this paper and outlines possible future works for improving the weighted multi-path C block.

\section{The Collection Block} \label{sec:collection}

The C building block aggregates values held by different devices through a spatial region into a single value in a selected device, by repeated application of an associative and commutative aggregation operator, where each iteration is executed in asynchronous rounds by every device in the network: partially aggregated values flow towards the selected source, while avoiding multiple aggregation of the same values. This two-faceted prerequisite, of \emph{acyclic} flows \emph{directed} towards the source, is met by relying on a given \emph{potential field}, approximating a certain measure of distance from the selected source. As long as information flows descending the potential field, cyclic dependencies are prevented and eventual reaching of the source is guaranteed. The C block thus computes a ``summary'' of given values $v_\deviceId$ in the local minima of a potential field $P(\deviceId)$, through a binary aggregating operator $\oplus$. Potential descent is enforced by splitting neighborhoods according to their potential value, obtaining the two disjoint sets:
\begin{align*}
D^-_\deviceId &= \bp{ \deviceId' \text{ linked to } \deviceId \text{ such that } P(\deviceId') < P(\deviceId) } \\
D^+_\deviceId &= \bp{ \deviceId' \text{ linked to } \deviceId \text{ such that } P(\deviceId') > P(\deviceId) }
\end{align*}

Two different implementations of the collection block have been proposed so far: \emph{single-path} and \emph{multi-path}. The single-path strategy $C_\text{sp}$ ensures that information flows through a forest in the network, by sending the whole partial aggregate of a device $\deviceId$ to the single device $m(\deviceId)$ with \emph{minimal} potential among devices connected to $\deviceId$. This is accomplished by repeatedly applying the following update rule:
\[
C_\text{sp}(\deviceId) = v_\deviceId \oplus \bigoplus_{\deviceId' \in D^+_\deviceId \wedge m(\deviceId') = \deviceId} C_\text{sp}(\deviceId'),
\]
which computes the partial aggregate in $\deviceId$ by combining together the value in $\deviceId$ and the partial aggregates in devices with higher potential for which $\deviceId$ is the selected output device $m(\deviceId')$.

The multi-path strategy $C_\text{mp}$, instead, allows information to flow through every path compatible with the given potential field. The partial aggregate of a device $\deviceId$ is thus divided equally among \emph{every} device $\deviceId'$ connected to $\deviceId$ with lower potential, by iteratively applying the following update rule:
\[
C_\text{mp}(\deviceId) = v_\deviceId \oplus \bigoplus_{\deviceId' \in D^+_\deviceId} \bp{C_\text{mp}(\deviceId') \oslash |D^-_{\deviceId'}| }; 
\]
where $\oslash$ is a binary operator $v \oslash n$ ``extracting the $n$-th root'', i.e., an element which aggregated with itself $n$ times produces the original value $v$.
Since information needs to be ``divisible'' for $\oslash$ to exist, \emph{multi-path} has a narrower scope than \emph{single-path}. However, both arithmetic-like operations ($+, \times, \ldots$) and idempotent operations ($\min, \max, \ldots$) are \emph{divisible} (through resp.~division, root extraction, identity); so that the situations where \emph{multi-path} is not applicable are rare in practice.

\section{Weighted Multi-path C} \label{sec:weighted}

The \emph{weighted multi-path} C develops on the multi-path strategy, by allowing partial aggregates to be divided unequally among neighbors. \emph{Weights} corresponding to neighbors are calculated in order to penalize devices which are likely to lose their ``receiving'' status, situation which can happen in two cases:
\begin{enumerate}
	\item if the ``receiving'' device is too close to the edge of the communication range of the ``sending'' device, it might step outside of it in the immediate future breaking the connection;
	
	\item if the potential of the ``receiving'' device is too close to the potential of the ``sending'' device, their relative role of sender/receiver might be switched in the immediate future, possibly creating an ``information loop'' between the two devices.
\end{enumerate}
\pagebreak
We can address both of these situations with a natural weight function $w(\deviceId,\deviceId')$, measuring how much of the information from $\deviceId$ should flow to $\deviceId'$, as the product of two corresponding factors $d(\deviceId,\deviceId') \cdot p(\deviceId,\deviceId')$:
\[
	d(\deviceId,\deviceId') =  R - D(\deviceId,\deviceId'), \qquad p(\deviceId,\deviceId') = \vp{P(\deviceId)-P(\deviceId')}
\]
where $R$ is the communication radius, $D(\deviceId,\deviceId')$ is the physical distance between devices $\deviceId, \deviceId'$ and $P(\deviceId)$ is the potential of device $\deviceId$. Notice that $w$ is positive and symmetric, hence it can be interpreted as an attribute of connection links representing both the amount of information to ``send'' and to ``receive''.
Since the $w(\deviceId, \deviceId')$ values do not sum up to any particular value, they need to be normalized by the factor $N(\deviceId) = \sum_{\deviceId' \in D^-_\deviceId} w(\deviceId, \deviceId')$ before use, obtaining normalized weights $w(\deviceId,\deviceId') / N(\deviceId')$.
The partial aggregates accumulated by devices can then be calculated as in $C_\text{mp}$ with the addition of weights, by iteratively applying the following update rule:
\[
C_\text{wmp}(\deviceId) = v_\deviceId \oplus \bigoplus_{\deviceId' \in D^+_\deviceId} \bp{C_\text{wmp}(\deviceId') \otimes \frac{w(\deviceId', \deviceId)}{N(\deviceId')} }; 
\]
where $\otimes$ is a binary operator $v \otimes k$ ``extracting'' a certain percentage $k$ of a local value $v$. 

\begin{figure}[t]
	\centering
	\includegraphics[width=0.32\textwidth]{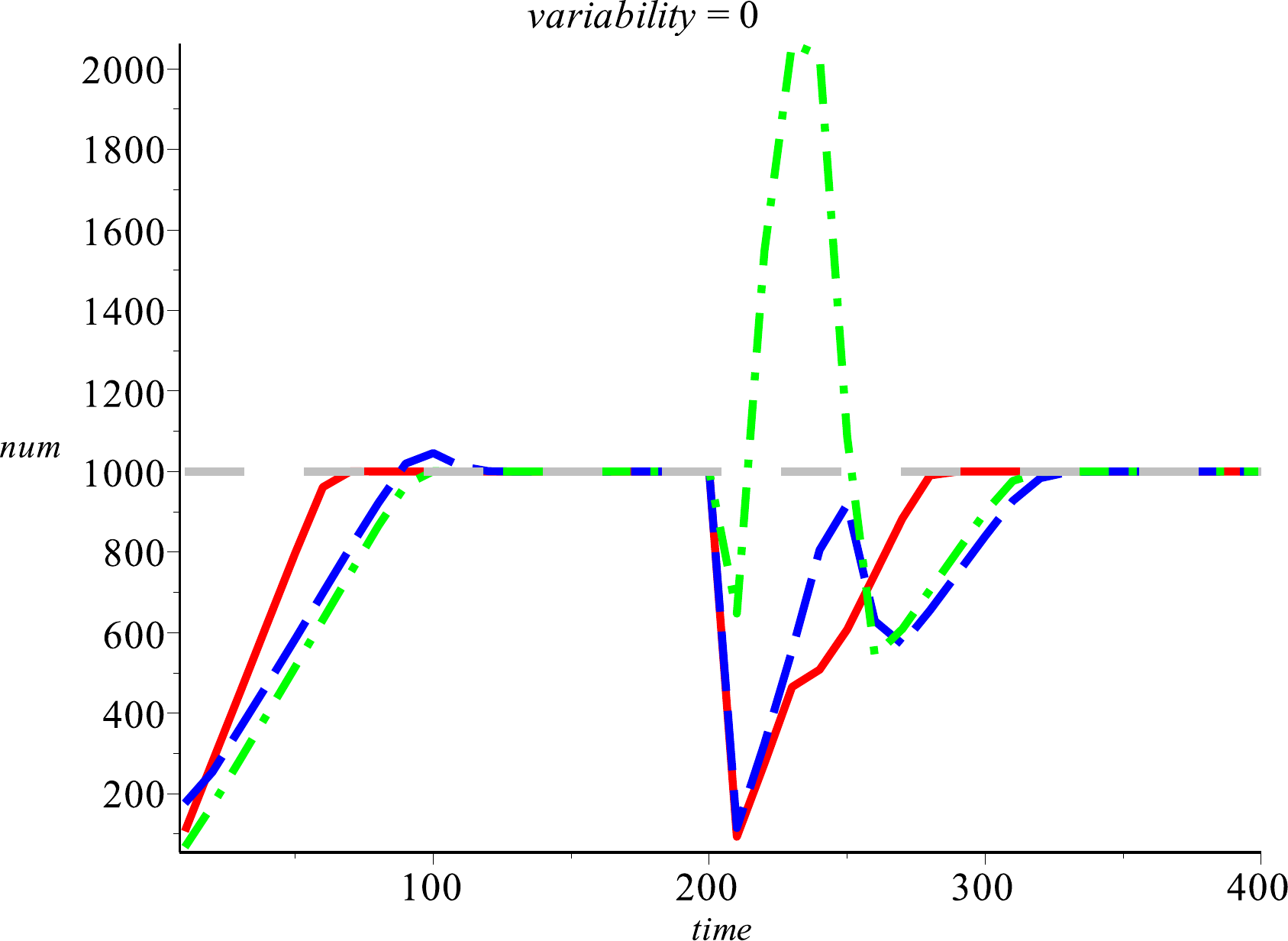}
	\includegraphics[width=0.32\textwidth]{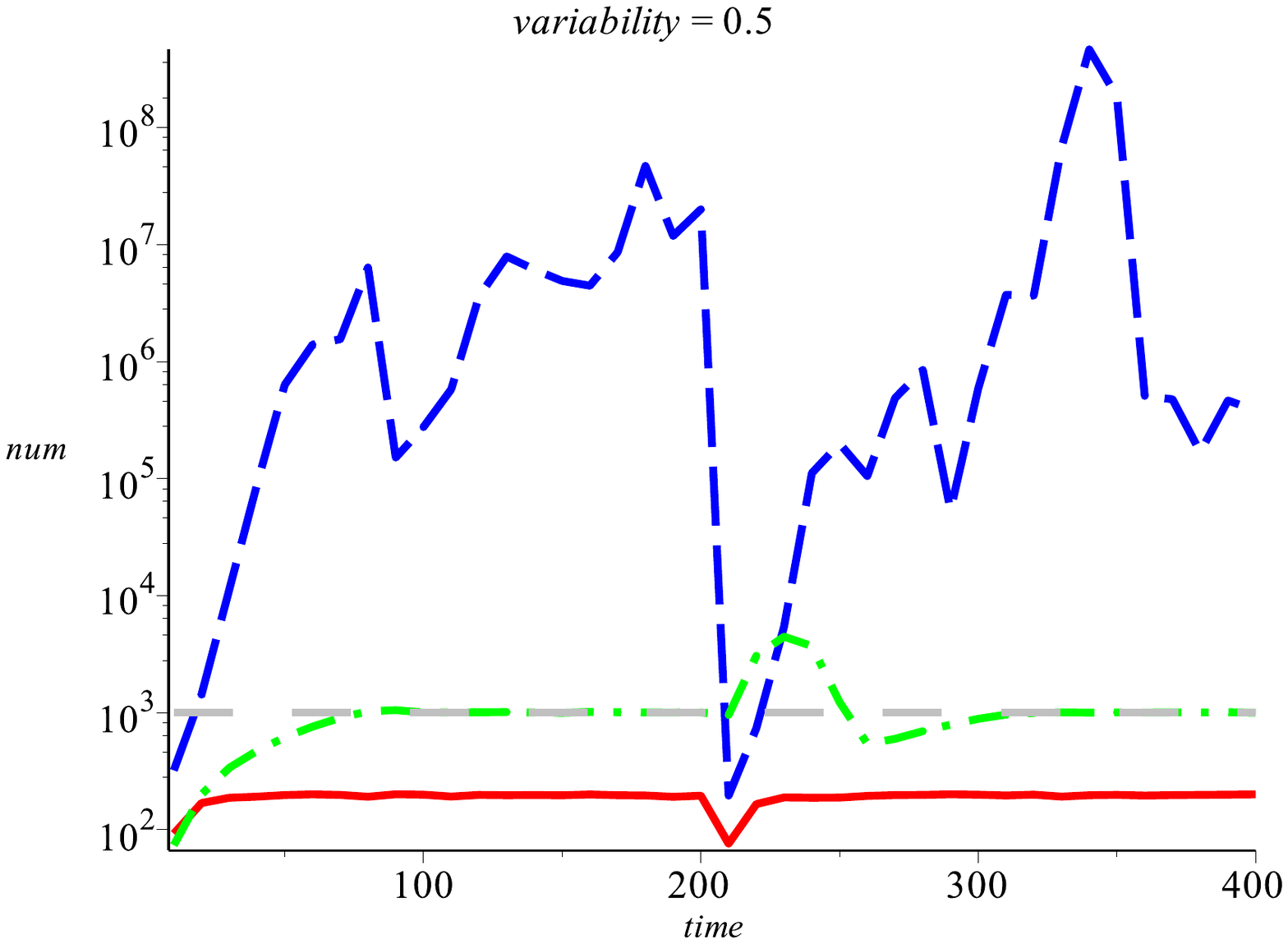}
	\includegraphics[width=0.32\textwidth]{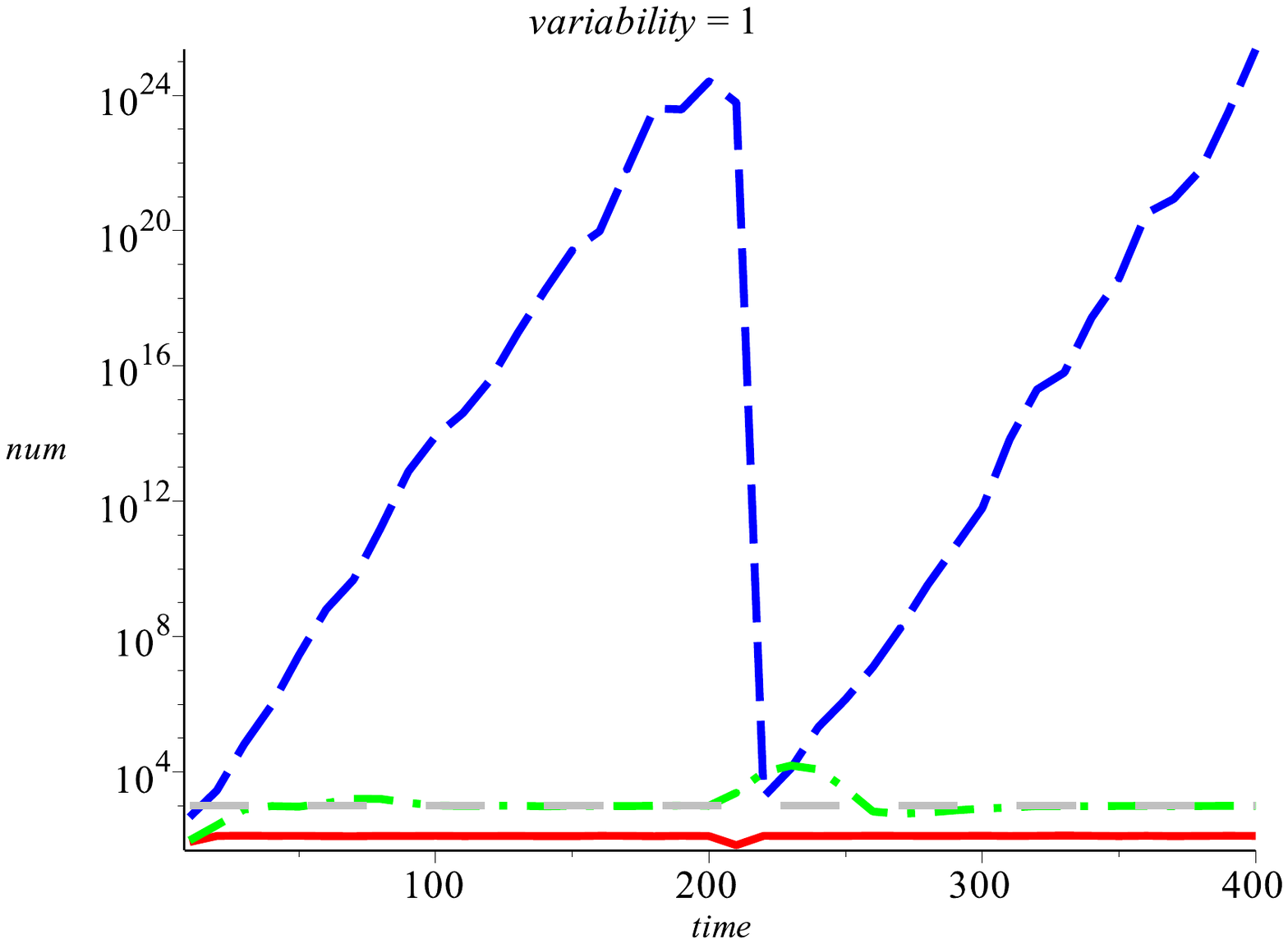} \\
	\includegraphics[width=0.32\textwidth]{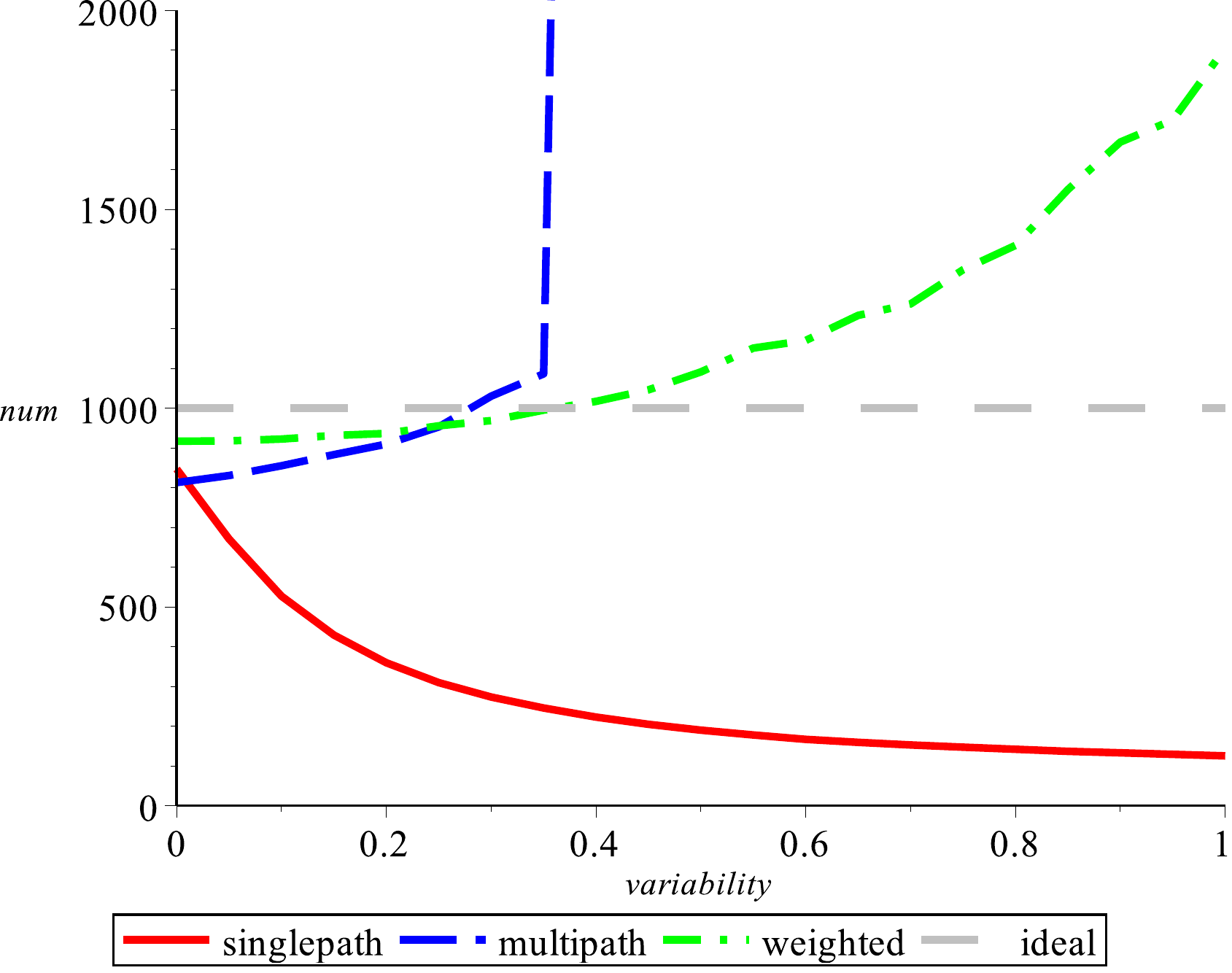}
	\vspace{-5pt}
	\caption{Number of devices measured through implementations of C, upon increasing variability with source change at $T = 200s$. Evolution through time is presented for variabilities $0$, $0.5$, $1$ (top), together with the average number in time window $0-400s$ for $21$ variabilities from $0$ to $1$ (bottom).} \label{fig:performance}
\end{figure}

\section{Evaluation} \label{sec:evaluation}

We compared \emph{weighted}, \emph{multi-path} and \emph{single-path} in an archetypal scenario of $1000$ devices randomly distributed along a $200m \times 20m$ corridor, with a $1s$ average computation rate and a $10m$ communication range. The source device was located on the right end of the corridor, then discontinuously moved to the left end at time $T = 200s$. The potential field was computed through BIS gradient \cite{bisgradient}, and the aggregation chosen to count the total number of devices (thus setting $\oplus=+$, $\otimes=\times$ and $v=1$ for each device). We tested degrees of variability ranging from $0$ (no movement, regular computation rate) to $1$ (short- and long-range movements, irregular computation rate in each device and between different devices), thus testing the algorithm in a scenario of increasing variability aimed at reproducing the worst possible case.

Figure~\ref{fig:performance} summarises the evaluation results; obtained with Protelis \cite{Protelis15} as programming language, Alchemist as simulator \cite{PianiniJOS2013} and the supercomputer OCCAM~\cite{16:occam:chep} as platform\footnote{For the sake of reproducibility, the actual experiment is made available at \url{https://bitbucket.org/gaudrito/audrito-bergamini-summarising}}. We run $100$ instances of each scenario and averaged the results, which had a significant relative standard error between them for high values of variability.
The tests showed that \emph{single-path} systematically underestimates the ideal value, with a similarly poor performance under variability $0.5$ and $1$ showing that accurate values are attainable only for low-variability scenarios.
Conversely, \emph{multi-path} overestimates the value with an exponentially-growing behavior (randomly ``reset'' from time to time) which gets exponentially worse with increased variability.
On the other hand, \emph{weighted multi-path} achieves an adequate accuracy even in highly volatile scenarios, scoring the best results under steady inputs for every value of variability. The recovery speed from input discontinuities is also comparable with that of the other algorithms, however, the resulting transients contain value peaks that are sometimes higher than those of other algorithms.\\[-20pt]

\section{Conclusion} \label{sec:conclusion}

We introduced \emph{weighted multi-path C}, a new algorithm for summarising distributed data improving over state-of-the-art implementations of the C block. Experimental evaluation shows that \emph{weighted} achieves adequate accuracy even in high-variability scenarios where other approaches are infeasible.
However, the new algorithm suffers from high (though short) error peaks in response to input discontinuities: in these situations, potential-descending data flows create loops leading to exponential increases in error. Future works could address this problem, either through detecting input discontinuities for triggering corrective actions, or by preventing the creation of loops through time-driven potential fields.\\[-20pt]

\bibliographystyle{eptcs}
\bibliography{long}
\end{document}

%% file: macros.tex
\newcommand{\FORGET}[1]{}

\newcommand{\deviceId}{\delta}

\definecolor{dark-gray}{gray}{0}
\newcommand{\bp}[1]{\left\lbrace #1 \right\rbrace}
\newcommand{\vp}[1]{\left\lvert #1 \right\rvert}